\pdfoutput=1
\documentclass[preprint,superscriptaddress]{revtex4}

\usepackage{graphicx}
\usepackage{bm}

\begin{document}

\title{Strength of the Spin-Fluctuation-Mediated Pairing Interaction in a High-Temperature Superconductor}

\author{T. Dahm}
\affiliation{Institute for Theoretical Physics, University of T\"ubingen, D-72076 T\"ubingen, Germany}

\author{V. Hinkov}
\affiliation{Max-Planck-Institute for Solid State Research, D-70569 Stuttgart, Germany}

\author{S. V. Borisenko}
\author{A. A. Kordyuk}
\author{V. B. Zabolotnyy}
\affiliation{IFW Dresden, P.O. Box 270116, D-01171 Dresden, Germany}

\author{J. Fink}
\affiliation{IFW Dresden, P.O. Box 270116, D-01171 Dresden, Germany}
\affiliation{BESSY, D-12489 Berlin, Germany}

\author{B. B{\"u}chner}
\affiliation{IFW Dresden, P.O. Box 270116, D-01171 Dresden, Germany}

\author{D. J. Scalapino}
\affiliation{Department of Physics, University of California, Santa Barbara, CA 93106-9530, USA}

\author{W. Hanke}
\affiliation{Institute for Theoretical Physics, University of W\"urzburg, D-97074 W\"urzburg, Germany}

\author{B. Keimer}
\affiliation{Max-Planck-Institute for Solid State Research, D-70569 Stuttgart, Germany}

\maketitle

\baselineskip24pt {\bf Theories based on the coupling between spin
fluctuations and fermionic quasiparticles are among the leading
contenders to explain the origin of high-temperature
superconductivity, but estimates of the strength of this interaction
differ widely \cite{eschrig_review}. Here we analyze the charge- and
spin-excitation spectra determined by angle-resolved photoemission
and inelastic neutron scattering, respectively, on the same crystals
of the high-temperature superconductor YBa$_2$Cu$_3$O$_{6.6}$. We
show that a self-consistent description of both spectra can be
obtained by adjusting a single parameter, the spin-fermion coupling
constant. In particular, we find a quantitative link between two
spectral features that have been established as universal for the
cuprates, namely high-energy spin excitations
\cite{hayden,tranquada,hinkov2,vignolle,lipscombe,fauque} and
``kinks" in the fermionic band dispersions along the nodal direction
\cite{damascelli,lanzara,kordyuk,borisenko,zabolotnyy}. The
superconducting transition temperature computed with this coupling
constant exceeds 150 K, demonstrating that spin fluctuations have
sufficient strength to mediate high-temperature superconductivity.}

Looking back at conventional superconductors, the most convincing
demonstration of the electron-phonon interaction as the source of
electron pairing was based on the quantitative correspondence
between features in the electronic tunneling conductance and the
phonon spectrum measured by inelastic neutron scattering (INS)
\cite{McMill,Parks}. The rigorous comparison of fermionic and
bosonic spectra was made possible by the Eliashberg theory
\cite{Eliashberg} which allowed one to derive the tunneling
conductance from the experimentally determined phonon spectrum.
Various difficulties have impeded a similar approach to the origin
of high-temperature superconductivity. First, the $d$-wave pairing
state found in these materials implies a strongly momentum-dependent
pairing interaction. A more elaborate analysis based on data from
momentum-resolved experimental techniques such as INS and
angle-resolved photoemission spectroscopy (ARPES)
is thus required. These methods, in turn, impose conflicting
constraints on the materials. In order to avoid surface-related
problems, most ARPES experiments have been performed on the
electrically neutral BiO cleavage plane in
Bi$_2$Sr$_2$Ca$_{n-1}$Cu$_{n}$O$_{2(n+2)+\delta}$ \cite{damascelli}.
However, as a consequence of electronic inhomogeneity this family of
materials exhibits broad INS spectra that greatly complicate a
quantitative comparison with ARPES data \cite{fauque}. Conversely,
compounds with sharp spin excitations, including
YBa$_2$Cu$_3$O$_{6+x}$, have generated problematic ARPES
spectra due to polar surfaces with charge distributions different from the bulk
\cite{damascelli}. Finally, an analytically rigorous
treatment of the spin-fluctuation-mediated pairing interaction is
difficult, because small expansion parameters used in the
traditional Eliashberg theory (such as the ratio of Debye and Fermi
energies) are missing \cite{huang}. Because of these difficulties,
widely different values have been quoted for the spin-fermion
coupling constant \cite{eschrig_review}.

The analysis of YBa$_2$Cu$_3$O$_{6.6}$ data reported here was made
possible by recent advances on several fronts. First, INS
experiments on this material now consistently yield high-quality
spin excitation spectra over a wide energy and momentum range
\cite{hayden,hinkov2}. Second, recent ARPES experiments on
YBa$_2$Cu$_3$O$_{6+x}$ \cite{borisenko,zabolotnyy} have overcome
problems related to polar surfaces and allowed the observation of
superconducting gaps and band renormalization effects (``kinks")
akin to those previously reported in La- and Bi-based cuprates
\cite{damascelli}. Third, calculations based on the two-dimensional
(2D) Hubbard model have demonstrated Fermi surfaces, single-particle
spectral weights, antiferromagnetic spin correlations, and
$d_{x^2-y^2}$ pairing correlations in qualitative agreement with
experimental measurements \cite{senechal,dahm,aichhorn1,maier}.
Numerically accurate solutions of this model can thus serve as a
valuable guideline for a treatment of the spin fluctuation
interaction in the cuprates. This is the approach we take here.

Recent quantum Monte Carlo calculations of the 2D Hubbard model
within the dynamical cluster approximation (DCA-QMC) \cite{maier}
have shown that the effective pairing interaction can be
parameterized in terms of the numerically computed spin
susceptibility $\chi(\vec{Q},\Omega)$ in the form
\begin{equation}
 V_{eff}(\vec{Q},\Omega)= \frac{3}{2}\ \bar U^2  \, \chi(\vec{Q},\Omega),
\label{eq:1}
\end{equation}
where $\bar U$ is the coupling strength, and that this interaction
generates reasonable values for the superconducting transition
temperature $T_c$. Here we follow a similar strategy, but use
$\chi(\vec{Q},\Omega)$ determined by INS, on high-quality detwinned
YBa$_2$Cu$_3$O$_{6.6}$ single crystals described previously
\cite{hinkov2}. In order to serve as input for the numerical
calculations, we have used an analytic form of ${\rm Im}\chi$ that
provides an excellent description of the INS data (see the
Supplementary Information). Fig. 1 shows a plot of this form in
absolute units. In the superconducting state, the spin excitations
exhibit the well-known ``hour glass'' dispersion, with a neck at the
wave vector $\vec{Q}=(\pi,\pi)$ characteristic of antiferromagnetism
in the copper oxide planes and the ``resonance'' energy $\Omega =
38.5$ meV. (We use a notation in which the lattice parameter $a$ and
the Planck constant $\hbar$ are set to unity.) The lower branch of
the hour glass appears to be influenced by materials-specific
details. For instance, recent INS work on La$_{2-x}$Sr$_x$CuO$_4$
indicates two characteristic energies \cite{vignolle,lipscombe},
rather than the single resonance found in YBa$_2$Cu$_3$O$_{6+x}$.
The upper branch of high-energy spin excitations, on the other hand,
is common to all copper oxides thus far investigated by INS in this
energy range
\cite{hayden,tranquada,bourges,hinkov2,vignolle,lipscombe,fauque}.
Moreover, while the resonance in YBa$_2$Cu$_3$O$_{6.6}$ disappears
above $T_c$ \cite{hinkov2}, the intensity of the spin excitations
above $\Omega \sim 50$ meV is not noticeably affected by the
superconducting transition and only decreases slowly upon further
heating \cite{hayden,hinkov2}.

We extract the second parameter in Eq. 1, the coupling strength
$\bar U$, from a combined analysis of the INS data parameterized in
this way and the fermionic band dispersions observed by ARPES on the
same crystals (see Methods). As noted before, bonding and
antibonding combinations of electronic states on the two Cu-O layers
in the YBa$_2$Cu$_3$O$_{6.6}$ unit cell give rise to two distinct
Fermi surfaces (Fig. 2). The most prominent signature of many-body
effects in the ARPES data, namely the ``kink" along the nodal
direction (cut \#1 in Figs. 3), is highlighted in Fig.~4, where the
bonding band is singled out by a proper choice of excitation energy.

We now proceed to a quantitative analysis of the renormalization of
the nodal band dispersion by spin fluctuations. Before describing
the results, we take a look at the kinematics of spin fluctuation
scattering near the nodal points, where complications from the
superconducting gap are absent. The spin fluctuations shown in Fig.
1 scatter electrons between bonding and antibonding bands, as
indicated by factors in the INS \cite{hayden,hinkov2} and ARPES
\cite{borisenko,zabolotnyy} cross sections. (Weak high-energy
excitations corresponding to intraband scattering \cite{pailhes2}
are neglected here.) An analysis of our numerical results below
shows that the scattering probability for electrons near the nodal
points is greatly enhanced when energy-momentum conservation allows
interband scattering into opposite nodal regions (green arrow in
Fig. 2). A look at the INS data (green line in Fig. 1) reveals that
this condition is satisfied by spin fluctuations of energy $\sim 80$
meV on the upper, universal, weakly temperature dependent branch of
the hour glass. At this characteristic energy in the temperature
range studied here, we therefore expect a weakly temperature
dependent anomaly in the band dispersion, as experimentally
observed.

A self-consistent numerical procedure with a single adjustable
parameter, the coupling strength $\bar U$, was developed to
quantitatively assess the influence of the spin-fluctuation
interaction on the spectral function determined by ARPES (see the
Methods section). Fig. 4 shows that an excellent description of the
nodal band dispersion over a wide energy range is obtained with
$\bar U=1.59$ eV, in rough agreement with values found in earlier
calculations based on phenomenological models of the spin
susceptibility \cite{eschrig_review}. In particular, both
theoretical and experimental results show deviations from linear
behavior (``kink") for $\omega \geq 80$ meV (arrow in Fig. 4). The
corresponding mass renormalization at the nodal point is ${\rm Re}\
Z_{A}=3.7$.

Fig. 3 shows a comparison of the calculated spectral weight
to the ARPES intensity for
all three cuts in Fig. 2. It is evident that the calculation yields
an excellent description of the ARPES data set over the entire
Brillouin zone without further fitting parameters. In particular,
the low-intensity region (``dip'') below the renormalized band in
cut \#3 can be understood as a consequence of coupling to the
magnetic resonance at the neck of the hour-glass dispersion. As
noted before \cite{eschrig_review,eschrig,chubukov}, the resonance
wave vector (red lines in Fig. 1 and 2) connects antinodal regions
in bonding and antibonding bands, and the resonance and gap energies
add up to the dip energy $\sim 65$ meV. The only noticeable
difference between the numerical and experimental data is the width
of the momentum distribution curves, which is substantially larger
in the ARPES data, presumably at least in part due to residual
surface inhomogeneities \cite{graser}.

Encouraged by the self-consistent description of INS and ARPES data,
we proceed to a calculation of the critical temperature of the
$d$-wave superconducting state arising from the exchange of spin
fluctuations. A recent DCA-QMC study has shown that a good estimate
of $T_c$ can be obtained by using the same effective interaction as
in the calculation of the self-energy \cite{maier}. For the set of
parameters found above, the linearized gap equations (see Methods)
yield the $d$-wave eigenvalue $\lambda_d=1.39$ in the normal state
($T=70$ K), corresponding to a transition temperature $T_c = 174$ K.
In principle, the INS and ARPES spectra would now have to be
remeasured at this higher temperature, the calculation repeated,
etc., until self-consistency is achieved. However, as the spectral
weight rearrangement of spin excitations in this temperature range
is largely confined to low excitation energies, our estimates for
$\bar U$ and $T_c$ are not expected to change substantially (see
Methods). In this context it is instructive to compare the
eigenvalue at $T=70$ K with the one obtained from the INS spectrum
at 5 K that includes the ``resonance'', $\lambda_d=1.49$. The
enhanced eigenvalue implies that the redistribution of spectral
weight of the spin excitations below $T_c$ leads to an increase of
the effective pairing strength. This lends support to an
interpretation of the magnetic resonance and associated antinodal
dip in terms of a superconductivity-induced feedback effect on the
spin fluctuation spectrum \cite{eschrig_review}. It is also
consistent with the large $2 \Delta_0 / k_B T_c$ ratio.

In summary, we have shown that data from two momentum-resolved
experimental probes of a cuprate superconductor can be related in a
quantitative fashion, in close analogy to the traditional analysis
of the electron-phonon interaction in conventional superconductors.
Our analysis is in overall agreement with conclusions drawn from
prior work based on phenomenological spin excitation spectra and/or
data from probes without momentum resolution
\cite{eschrig_review,eschrig,chubukov,carbotte,timusk1,timusk2}, and
it resolves some problems that appeared in the context of these
studies. In particular, models that attribute the nodal kink in
ARPES either directly to the magnetic resonance
\cite{manske,eschrig} or to incoherent scattering processes from a
node into gapped states at the antinode \cite{chubukov} generally
predict that the kink is strongly modified by the onset of
superconductivity, whereas the experiments indicate at most a weak
effect at $T_c$ \cite{kordyuk}. The node-to-node interband
scattering process mediated by weakly temperature dependent,
universal, incommensurate spin excitations we have identified
provides a straightforward explanation of this observation. As the
incommensurate spin excitations persist into the optimally doped
\cite{vignolle,fauque} and overdoped \cite{lipscombe} regimes, it
also explains the persistence of both the kink and superconductivity
at high doping, where feedback effects related to the magnetic
resonance are progressively reduced \cite{timusk1}. There is thus no
need to invoke phonon scattering at this level \cite{lanzara}. While
some contribution of phonons to the nodal kink cannot be ruled out,
recent work has shown that it is hard to obtain a quantitative
description of the kink based on the electron-phonon interaction
alone \cite{chubukov,giustino,heid}.

It was previously shown \cite{woo} that the change in the magnetic
exchange energy between the normal and superconducting states is
more than enough to account for the cuprate superconducting
condensation energy. However, the crucial question whether the
exchange of magnetic spin-fluctuations actually has the strength to
give rise to high-$T_c$ pairing, was not answered. Here we have
shown that this interaction can generate $d$-wave superconducting
states with transition temperatures comparable to the maximum $T_c$
observed in the cuprates. In any given material, especially
underdoped cuprates such as YBa$_2$Cu$_3$O$_{6.6}$, a variety of
effects not considered in our analysis can reduce the actual $T_c$,
including vertex corrections of the spin fluctuation interaction
\cite{huang}, phase fluctuations of the order parameter, competition
with other types of order, and pair breaking by phonons and
impurities. It is also possible that phonons
\cite{lanzara,giustino,heid} or higher-energy excitations
\cite{yazdani} contribute to the pairing interaction. However, our
analysis indicates that the exchange of spin excitations already
directly observed by INS is a major factor driving the high
temperature superconducting state in the cuprates.

\section*{Methods}

The ARPES measurements were performed on the same
YBa$_2$Cu$_3$O$_{6.6}$ crystals used for the INS experiment, thus
avoiding systematic uncertainties invariably associated with
measurements on different materials. The details of the ARPES
experiments have been described elsewhere
\cite{borisenko,zabolotnyy}. Usually, YBCO single crystals cleave
between the Ba-O and Cu-O chain layers, resulting in an effective
overdoping of the Cu-O layer closest to the surface
\cite{borisenko,zabolotnyy,hossain}. A recent comprehensive study
has revealed, however, that in some cases the ARPES spectra are
dominated by a signal from the nominally doped Cu-O plane
\cite{zabolotnyy}. Here we present data taken on a particular spot
on the surface after one such successful cleave (Figs. 2, 3). The
strong many-body renormalization of the band structure typical for
underdoped cuprates (Fig. 4) as well as the anisotropic
superconducting gap (cuts \#2, 3 in Fig. 3) in the ARPES spectra
demonstrate that contributions from the overdoped surface-related
component are negligible. The superconducting component we observe
corresponds to the nominal doping, as follows from the size of the
gap and the temperature evolution of the coherence peaks which
disappear above the bulk SC transition temperature
\cite{zabolotnyy}.

The self-consistent numerical calculation we have used is based on
the self-energy diagram displayed in Fig. 4. The Green's functions
$G(\vec{k},\omega)$ on antibonding ($A$) and bonding ($B$) bands can
be written as \cite{dahm}
\begin{equation}
  G_{A,B}(\vec{k},\omega)=\frac{\omega Z_{A,B}+\tilde\epsilon^{A,B}_k}{(\omega Z_{A,B})^2-
    \left(\tilde\epsilon^{A,B}_k\right)^2-({\rm Re}\ Z_{A,B}(\vec{k},\omega=0)\Delta_k)^2}
\end{equation}
where $\Delta_k$ is the superconducting gap, which we assume to be
of the $d$-wave form $\Delta_k=\Delta_0(\cos k_x-\cos k_y)/2$ with
$\Delta_0=30$~meV, and $\tilde\epsilon^{A,B}_k = \epsilon^{A,B}_k +
\xi_{A,B}$ is the renormalized band structure. The
unrenormalized band dispersions $\epsilon^{A,B}_k$ were derived from
tight-binding fits to the ARPES Fermi surface in combination with
additional information from band structure calculations (see the Supplementary Information).
We have found that the results of our
calculations are quite robust against modifications of
$\epsilon^{A,B}_k$ (see the Supplementary Information). Finally, $\omega
Z_{A,B}(\vec{k},\omega)=\omega-\frac{1}{2} \left(
\Sigma_{A,B}(\vec{k},\omega) - \Sigma_{A,B}^*(\vec{k},-\omega) \right)
+i\Gamma_{el}$ is the mass renormalization function and
$\xi_{A,B}(\vec{k},\omega)=\frac{1}{2} \left(
\Sigma_{A,B}(\vec{k},\omega) + \Sigma_{A,B}^*(\vec{k},-\omega) \right) $
the energy shift function. Apart from an elastic scattering rate
$\Gamma_{el}\sim 30$ meV, which accounts for impurity scattering,
the mass renormalization function is determined by the
imaginary part of the electron self-energy $\Sigma_{A,B}$, which can
be written as
\begin{equation}
  {\rm Im}\ \Sigma_{A,B}(\vec{k},\omega)=\frac{1}{\pi N}\sum_Q\int^\infty_{-\infty}
d\Omega[n(\Omega)+f(\Omega-\omega)]\ {\rm Im}\, V_{eff} (\vec{Q},\Omega)
\ {\rm Im}\ G_{B,A} (\vec{k}-\vec{Q},\omega-\Omega) \ .
\end{equation}
Here $\sum_Q$ denotes a sum of the in-plane momenta over the full
Brillouin zone, $n$ and $f$ are the Bose- and Fermi-functions, respectively, and
$V_{eff}$ is the spin fluctuation interaction (Eq. 1). The real parts of  $\Sigma_{A,B}$ that enter into Eq. 2
are obtained by Kramers-Kronig transformations.
Note that the self-energy in the antibonding band is
determined by the interaction with the bonding band and vice versa.

Together with Eq. 1, this defines a self-consistent system of equations with a single
adjustable parameter, the coupling strength $\bar U$. Starting with
noninteracting values for the Green's functions, these equations were
solved iteratively until convergence is achieved. The renormalized band
dispersion and spectral weight, $f(\omega)\, {\rm Im}\, G(\vec{k},\omega)$,
can then be compared to ARPES data.

The linearized gap equations
\begin{eqnarray}
  \lambda_d\ {\rm Im}\ \phi_{A,B}(\vec{k},\omega)&=&
\frac{1}{\pi N}\sum_{k'}\int^\infty_{-\infty}
    d\omega'[n(\omega-\omega')+f(-\omega')] \\ \nonumber
&& {\rm Im}\, V_{eff}(\vec{k}-\vec{k}',\omega-\omega')\ {\rm Im}
    \left\{\frac{\phi_{B,A}(\vec{k}',\omega')}{(\omega'Z_{B,A})^2-
(\tilde\epsilon^{B,A}_{k'})^2}\right\}
\label{eq:9}
\end{eqnarray}
were solved for the same set of parameters. Note that the INS data used
for the calculations were taken at $T=5$ and 70
K, while the ARPES data were taken at 30 K. Since the changes in
the superconducting gap and INS spectrum between 5 and 30 K are negligible,
we use the 5 K INS results along with the 30 K ARPES data to determine
the coupling constant $\bar U$. As $T$ is raised further, the superconducting
gap decreases, and there is a shift of ${\rm Im}\chi$ to lower frequencies.
However, we expect that $\bar U$ is unchanged for the range of temperatures
of interest, because it is determined by weakly $T$-dependent high-energy
processes.

For discussions of the influence of a high-energy cutoff in
$\chi(\vec{Q},\Omega)$ and of a normal-state pseudogap on the
results of the numerical calculations, see the Supplementary
Information.

\bibliographystyle{naturemag}
\bibliography{spinfluctuations}

\vspace{.1in}

\noindent \textbf{Acknowledgements.} This project is part of the
Forschergruppe FOR538 of the German Research Foundation. DJS
acknowledges the Center for Nanophase Material Sciences at Oak Ridge
National Laboratory, U.S. Department of Energy. We thank P. Bourges
and A. Ivanov for discussions.

\vspace{.1in}

\noindent Correspondence and requests for materials should be
addressed to B.K. (b.keimer@fkf.mpg.de).

\vspace{.1in}

\noindent Supplementary Information accompanies this paper on
www.nature.com/naturephysics.

\vspace{.1in}

\noindent \textbf{Competing financial interests.} The authors
declare that they have no competing financial interests.

\begin{figure}[htb]
\begin{center}
\includegraphics[width= \columnwidth, angle=0]{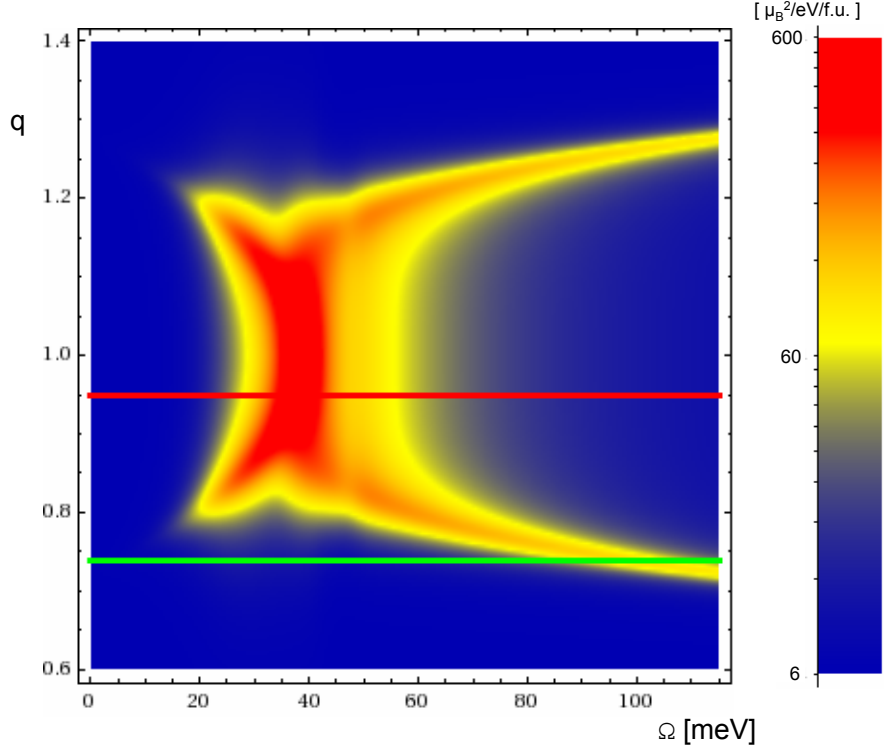}
\end{center}
\caption{Intensity of spin excitations along $\vec{Q}=q (\pi,\pi)$ resulting
from numerical fits to the INS spectra of YBa$_2$Cu$_3$O$_{6.6}$ at $T=5$ K. The maximum value corresponds to 600 $\mu_B^2/$eV/f.u. The green (red)
lines mark wave vectors connecting nodal (antinodal) regions on different
Fermi surfaces (Fig. 2).
}
\label{fig:nodaldispersion}
\end{figure}

\begin{figure}[htb]
\begin{center}
\includegraphics[width=0.6 \columnwidth, angle=270]{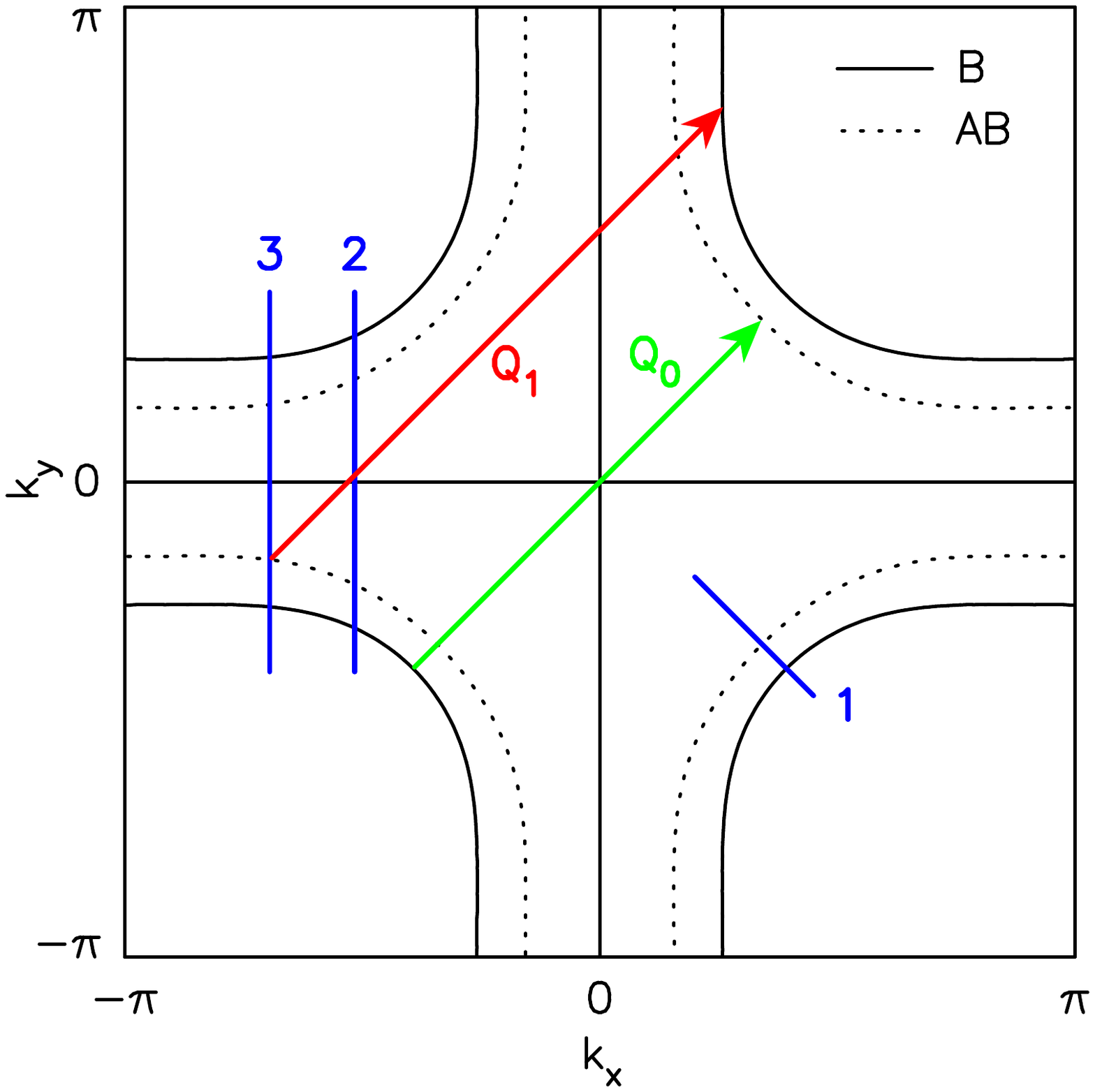}
\end{center}
\caption{Results of tight-binding fits of the Fermi surfaces determined by ARPES for the bonding (solid line) and antibonding (dotted line) bands.
The blue lines denote the three cuts along which experimental and theoretical spectral functions are compared (Fig. 3).
The green and red arrows indicates spin-fluctuation-mediated scattering processes discussed in the text. }

\end{figure}

\begin{figure}[htb]
\hspace*{-2.5cm}
\includegraphics[width=1.2\columnwidth, angle=0]{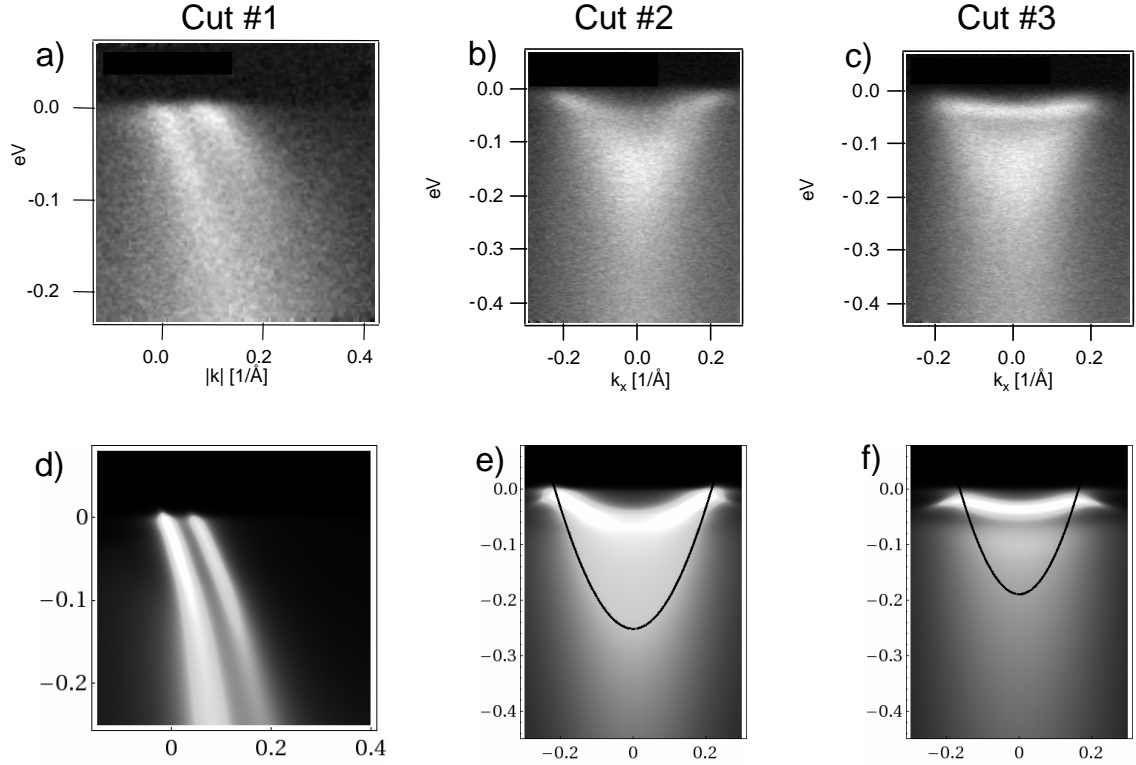}
\caption{{\bf a}--{\bf c} ARPES intensities  at 30 K along the three cuts shown in Fig. 2, obtained with photon energies 60 eV
(cut \#1) and 55 eV (cuts \#2, 3). {\bf d}--{\bf f} Theoretical spectral functions at $T=5$ K, multiplied by the Fermi function at 30 K (see Methods).
Intensities in bonding and antibonding bands in panel {\bf d} were multiplied by 0.75 and 0.25, respectively, to take account of matrix element effects.
The black lines in panels {\bf e} and {\bf f} show the unrenormalized antibonding band dispersion.}
\end{figure}

\begin{figure}[htb]
\begin{center}
\includegraphics[width=0.75 \columnwidth, angle=270]{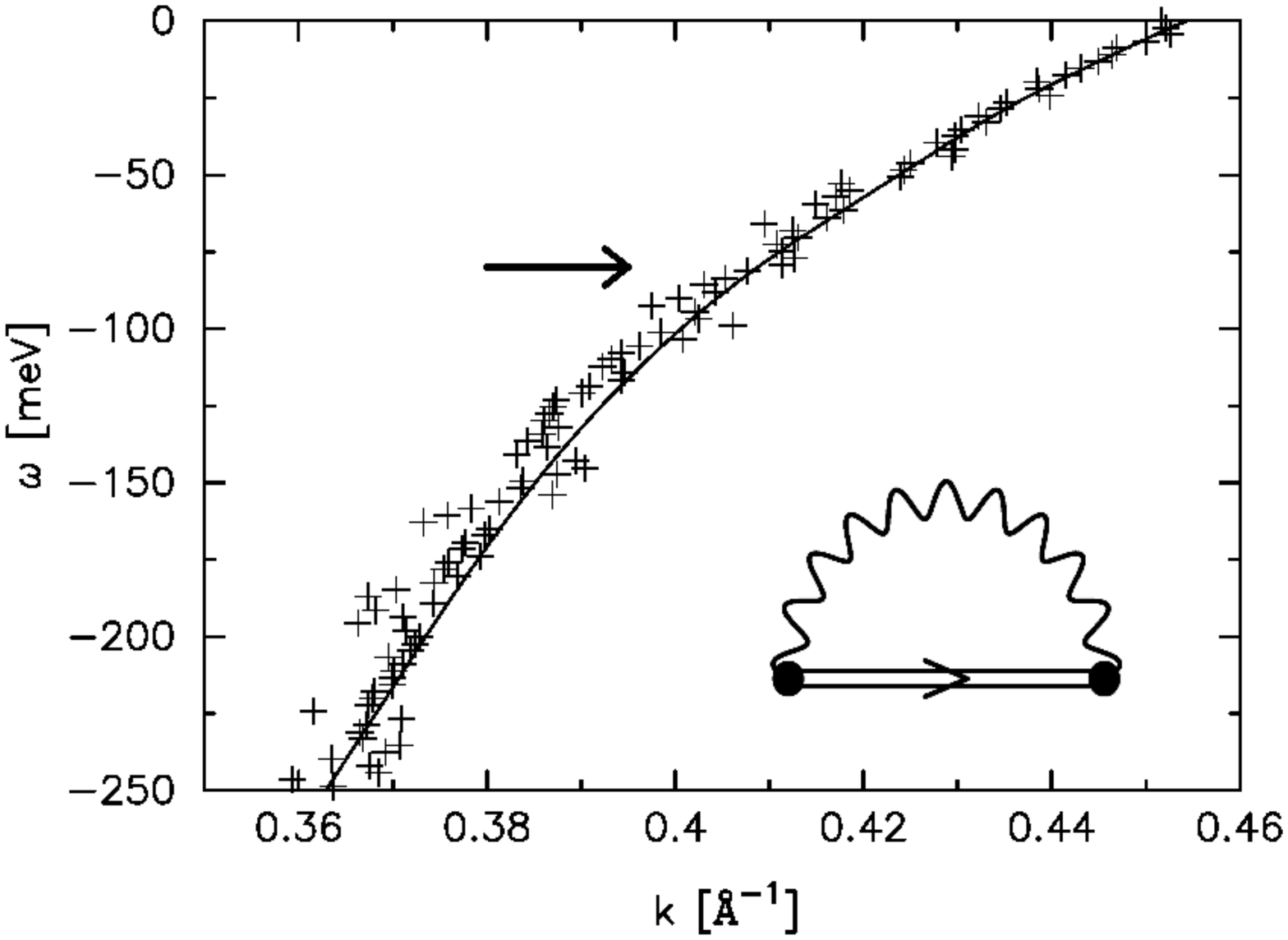}
\end{center}
\caption{Nodal dispersion of the ARPES data (crosses) for the bonding band compared with the same nodal dispersion of the
calculation (solid line). Here, the coupling constant $\bar{U}$ was adjusted such that the slopes of the curves (renormalized
Fermi velocity) match near zero energy. The inset shows a self-energy diagram for the exchange of spin fluctuations. The
wiggly line represents the interaction given by Eq.~1, and the straight double line is the self-consistent single-particle Green's function.}
\end{figure}

\clearpage

\newpage

\setcounter{figure}{0}
\setcounter{page}{1}

\vspace*{5cm}

\begin{center}
{\bf \large Supplementary Information for ``Strength of the Spin-Fluctuation-Mediated Pairing Interaction in a High-Temperature Superconductor''}
\end{center}

\newpage

\subsection*{Fit to INS data}

The INS results were fitted by simple analytic forms which could then be numerically integrated in the self-energy and $T_c$ calculations. In constructing these fits, the energy range was divided into upper ($u$) and lower ($\ell$) branches corresponding to energies above and below the resonance energy $\omega_r$.
\begin{equation}
  {\rm Im}\ \chi(\vec{Q},\Omega)={\rm Im}\ \chi_\ell(\vec{Q},\Omega)+{\rm Im}\ \chi_u(\vec{Q},\Omega)
\label{eq:A1}
\end{equation}

The data in the superconducting state at $T=5$ K were fitted as follows. On the lower branch
\begin{equation}
  {\rm Im}\ \chi_\ell(\vec{Q},\Omega)=N_0N_\ell(\Omega)N_L(\Omega)
    \frac{2(2\Omega_r-\Omega)\gamma_Q}{((2\Omega_r-\Omega)^2-\Omega^2_Q)^2+
    (2(2\Omega_r-\Omega)\gamma_Q)^2}
\label{eq:A2}
\end{equation}
Here the frequency $\Omega$ is measured in meV with the resonance
frequency $\Omega_r=38.5$ meV. The prefactors are given by
\begin{equation}
  N_\ell(\Omega)=1-\frac{1}{1+\exp[(\Omega-24)/3]}
\label{eq:A3}
\end{equation}
and
\begin{equation}
  N_L(\Omega)=\frac{1}{1+\exp[\Omega-42]}
\label{eq:A4}
\end{equation}
These factors essentially cut off the lower branch at frequencies above 42 meV and below 24 meV (spin gap). The momentum dispersion of the resonance mode is given by
\begin{equation}
  \Omega_Q=\Omega_r+s_a(Q_a-Q_{0a})^2+s_b(Q_b-Q_{0b})^2
\label{eq:A5}
\end{equation}
with ${\bf Q}_0=(Q_{0a},Q_{0b})$ the antiferromagnetic wave vector,
$s_a=280$ meV \AA$^2$ and $s_b=455$ meV \AA$^2$. The damping width
$\gamma_Q$ has an angular dependence of the form
\begin{equation}
  \gamma_Q=\gamma_a+\Delta\gamma\frac{(Q_a-Q_{0a})^2}{\left|{\bf Q}-{\bf Q}_0\right|^2}
\label{eq:A6}
\end{equation}
where the index $a$ denotes the $a$-axis direction in the Brillouin zone and $b$ the $b$-axis direction, respectively. The parameter values are $\gamma_a=1.5$ meV and $\Delta\gamma=6.5$ meV.

For the upper branch we have
\begin{equation}
  {\rm Im}\ \chi_u(\vec{Q},\Omega)=N_0N_{0u}N_u(\Omega)N_{dip}(\Omega)N_{rot}(\vec{Q})
    \frac{2\Omega\Gamma}{(\Omega^2-\Omega_Q^2)^2+(2\Omega\Gamma)^2}
\label{eq:A7}
\end{equation}
Here, $\Gamma=11$ meV and
\begin{equation}
  \Omega_Q=\Omega_r+\left[S^{2/p}_a(Q_a-Q_{0a})^2+S^{2/p}_b(Q_b-Q_{0b})^2\right]^{p/2}
\label{eq:A8}
\end{equation}
where $p=4$ is the power of the dispersion of the upper branch
(somewhat steeper than quadratic) with $S_a=4830$ meV \AA$^p$ and
$S_b=10065$ meV \AA$^p$. The prefactor $N_{0u}$ is 1.35 and
\begin{equation}
  N_u(\Omega)=1-\frac{1}{1+\exp[(\Omega-36)/1.5]}
\label{eq:A9}
\end{equation}
provides a cut-off of the upper branch for energies below 36 meV. The prefactor
\begin{equation}
  N_{dip}(\Omega)=1-0.2\exp\left[-\left(\Omega-\Omega_{dip}\right)^2/\left(2\sigma^2_{dip}\right)\right]
\label{eq:A10}
\end{equation}
with $\Omega_{dip}=47$ meV and $\sigma_{dip}=2$ meV accounts for the reduction
in intensity just above the resonance. The prefactor $N_{rot}$ has a momentum
dependence of the form
\begin{equation}
  N_{rot}(\vec{Q})=1-0.3\left(\frac{(Q_a-Q_{0a})^2-(Q_b-Q_{0b})^2}{\left|{\bf Q}-{\bf Q}_0\right|^2}\right)^2
\label{eq:A11}
\end{equation}
accounting for the 45 degree rotation of the signal above the resonance  [S1, S2]. The overall normalization factor $N_0$ is chosen such that the momentum integrated ${\rm Im}\ \chi(\vec{q},\Omega)$ has a peak value of $16\ \mu^2_B/{\rm eV}/{\rm f.u.}$ in the superconducting state at 5 K, which is known from previous work [S1, S3, S4]. In order to account for lattice symmetry, the function ${\rm Im}\ \chi(\vec{q},\Omega)$ is cut at the zone center and continued periodically.

Let us now discuss the fitting formula for the normal state data at 70 K. Again, the fitting formula is subdivided into an upper and lower branch. For the lower branch we have

\begin{eqnarray}
  {\rm Im}\ \chi_\ell(\vec{Q},\Omega) & = & N_0N_{0\ell}N_{lin}(\Omega)N_\ell(\Omega)N_L(\Omega) \\
    & & \{\exp[-4\ln2[((Q_a-Q_{1a})/\sigma_a)^2+((Q_b-Q_{1b})/\sigma_b)^2]] \nonumber \\
    & & +\exp[-4\ln2[((Q_a-Q_{2a})/\sigma_a)^2+((Q_b-Q_{2b})/\sigma_b)^2]]\} \nonumber
\label{eq:A12}
\end{eqnarray}
Here, the two Gaussians describe an energy-independent incommensurability with $\sigma_a=0.125$ r.l.u., $\sigma_b=0.17$ r.l.u., ${\bf Q}_1=(Q_{1a},Q_{1b})=
(0.575,0.5)$ r.l.u., and ${\bf Q}_2=(Q_{2a},Q_{2b})=(0.425,0.5)$ r.l.u. The
normalization factors are given by $N_{0\ell}=0.001807$,
\begin{equation}
  N_\ell(\Omega)=1-\frac{0.5}{1+\exp[(\Omega-29.5)/2]}
\label{eq:A13}
\end{equation}
and
\begin{equation}
  N_L(\Omega)=\frac{1}{1+\exp[(\Omega-37)/2]}.
\label{eq:A14}
\end{equation}
These factors essentially cut off the lower branch at frequencies above 37 meV and partially below 29.5 meV. The factor
\begin{equation}
N_{lin} \left( \Omega \right) = \left\{
\begin{array}{ccl}
1 & \mbox{for} & \Omega \ge 27 \, \mbox{meV} \\
\Omega / 27  & \mbox{for} & \Omega < 27 \, \mbox{meV}
\end{array}
\right.
\label{eq:A15}
\end{equation}
represents a simple linear decrease below 27 meV consistent with previous work [S5]. For the upper branch we have the same expression as in the superconducting state at 5 K except that
\begin{equation}
  N_{dip}(\Omega)=1
\label{eq:A16}
\end{equation}
and
\begin{equation}
  N_u(\Omega)=1-\frac{1.02}{1+\exp[(\Omega-37)/2]}.
\label{eq:A17}
\end{equation}

Figure S1 shows the momentum integrated ${\rm Im}\ \chi(\vec{q},\Omega)$ in absolute units obtained from the two formulae.  The ``sum-rule" integral
\begin{equation}
  S=\int^\infty_0\frac{d\Omega}{\pi}\left\langle\frac{{\rm Im}\ \chi(Q,\Omega)}{g^2\mu^2_B}\right\rangle_Q
\label{eq:A18}
\end{equation}
for the two fits gives $S=0.071/{\rm f.u.}$ for the normal phase  ($T=70$ K) and $0.070/{\rm f.u.}$ for the superconducting  phase ($T=5$ K). These numbers are in reasonable agreement with each other.

Figure S2 shows typical measured scans compared with our fit formula convoluted with the instrumental resolution function and demonstrates the good agreement.

In order to check the sensitivity of our results to the high energy part of the magnetic excitation spectrum,
we have repeated our calculations with a spectrum cut off at 200 meV. As a result we found that the coupling constant
$\bar{U}$ had to be increased by 3 percent. $T_c$ was found to decrease by 1 Kelvin, showing that the details of the
high energy part of the spectrum do not affect $T_c$ very much.

Further on, we have phenomenologically added a normal state $d$-wave pseudogap         of 30 meV into our calculation and found that $T_c$ increases by 20 percent, due to the suppression of pairbreaking low-energy magnetic excitations. We conclude that the influence of the pseudogap does not alter our two major findings: the high value of $T_c$ and the
nodal kink generated by the upper branch of the hourglass.

\subsection*{Fit to fermionic band dispersions}

As a starting point for our theoretical calculation we also need the unrenormalized dispersions $\epsilon^{A,B}_k$ for the bonding (B) and antibonding (A) bands of the two-layer system. In contrast to previous calculations, here we keep the renormalized Fermi surface fixed during the iterative solution of Eqs. (1--3) of the main manuscript, because the Fermi surface is known from the ARPES data. In order to achieve this, we keep the renormalized quantity $\tilde\epsilon^{A,B}_k=\epsilon^{A,B}_k+{\rm Re}\ \xi_{A,B}(k,\omega=0)$ fixed during the calculation and obtain it from tight-binding fits to the ARPES data of the form

\begin{equation}
\mu_{A,B}-2t_{A,B}(\cos k_x+\cos k_y)+4t'_{A,B}\cos k_x
    \cos k_y-2t''_{A,B}(\cos 2k_x+\cos 2k_y)
\label{eq:8}
\end{equation}

with the parameters

\[
\begin{array}{cccc}
\mu_A=556\ \mbox{meV}&t_A=409\ \mbox{meV}&t'_A=150\ \mbox{meV}&t''_A=40\ \mbox{meV}\\
\mu_B=417\ \mbox{meV}&t_B=550\ \mbox{meV}&t'_B=231\ \mbox{meV}&t''_B=67\ \mbox{meV}
\end{array}
\]

which yield excellent descriptions of the ARPES Fermi surfaces (Fig. 2). These parameters are scaled such that the unrenormalized Fermi velocity at the nodal point in the antibonding band equals 5 eV\AA, as found in {\it ab-initio} bandstructure calculations [S6].

In order to test the sensitivity of the numerical results to the assumed unrenormalized Fermi velocity, we have repeated the calculations with an unrenormalized Fermi velocity of 4 eV\AA, a value that deviates from the {\it ab-initio} predictions for YBa$_2$Cu$_3$O$_{6+x}$ but is close to the one found in Bi$_2$Sr$_2$Ca$_{1}$Cu$_{2}$O$_{8+\delta}$ [S7]. In order to reproduce the renormalized Fermi velocity, the coupling constant $\bar{U}$ has to be reduced from 1.59 to 1.23 eV, and the nodal mass renormalization drops from 3.7 to 3.0. As shown in Fig. S3, the agreement of the numerical and ARPES data for the nodal dispersion at higher energies is worse than that obtained with the more realistic parameters in the main text. Nonetheless, the estimate of the critical temperature for $d$-wave superconductivity remains high. For the INS spectrum at 70 K, we obtain a $\lambda_d=1.28$, corresponding to $T_c = 140$ K.

Correspondence and requests for materials should be addressed to B.K.
(b.keimer@fkf.mpg.de).

\subsection*{References}

\noindent S1. Hayden, S. M., {\it et al.} The structure of the high-energy spin excitations in a high-transition-temperature superconductor. {\it Nature} {\bf 429}, 531--534 (2004).

\noindent S2. Hinkov, V. {\it et al.} Spin dynamics in the pseudogap state of a high-temperature superconductor {\it Nat. Phys.} {\bf 3}, 780--785 (2007).

\noindent S3. Bourges, P. {\it et al.} High-energy spin excitations in {YBa$_2$Cu$_3$O$_{6.5}$}. {\it Phys. Rev. B} {\bf 56}, R11439 (1997).

\noindent S4. Fong, H. F. {\it et al.} Spin susceptibility in underdoped {YBa$_2$Cu$_3$O$_{6+x}$}. {\it Phys. Rev. B} {\bf 61}, 14773 (2000).

\noindent S5. Stock, C. {\it et al.} Dynamic stripes and resonance in the superconducting and normal phases of {YBa$_2$Cu$_3$O$_{6.5}$ ortho-II}. {\it Phys. Rev. B} {\bf 69}, 014502 (2004).

\noindent S6. Andersen, O. K., Liechtenstein, A. I., Jepsen, O., Paulsen, F. {LDA} energy bands, low-energy {hamiltonians}, $t'$, $t''$, $t\bot(k)$, and {$J\bot$}. {\it J.~Phys. Chem. Solids} {\bf 56}, 1573 (1995).

\noindent S7. Kordyuk, A. A. {\it et al.}  Bare electron dispersion from experiment: Self-consistent self-energy analysis of photoemission data. {\it et al.}, {\it Phys. Rev. B} {\bf
71}, 214513 (2005).

\begin{figure}[htb]
\begin{center}
\includegraphics[width=0.6 \columnwidth, angle=270]{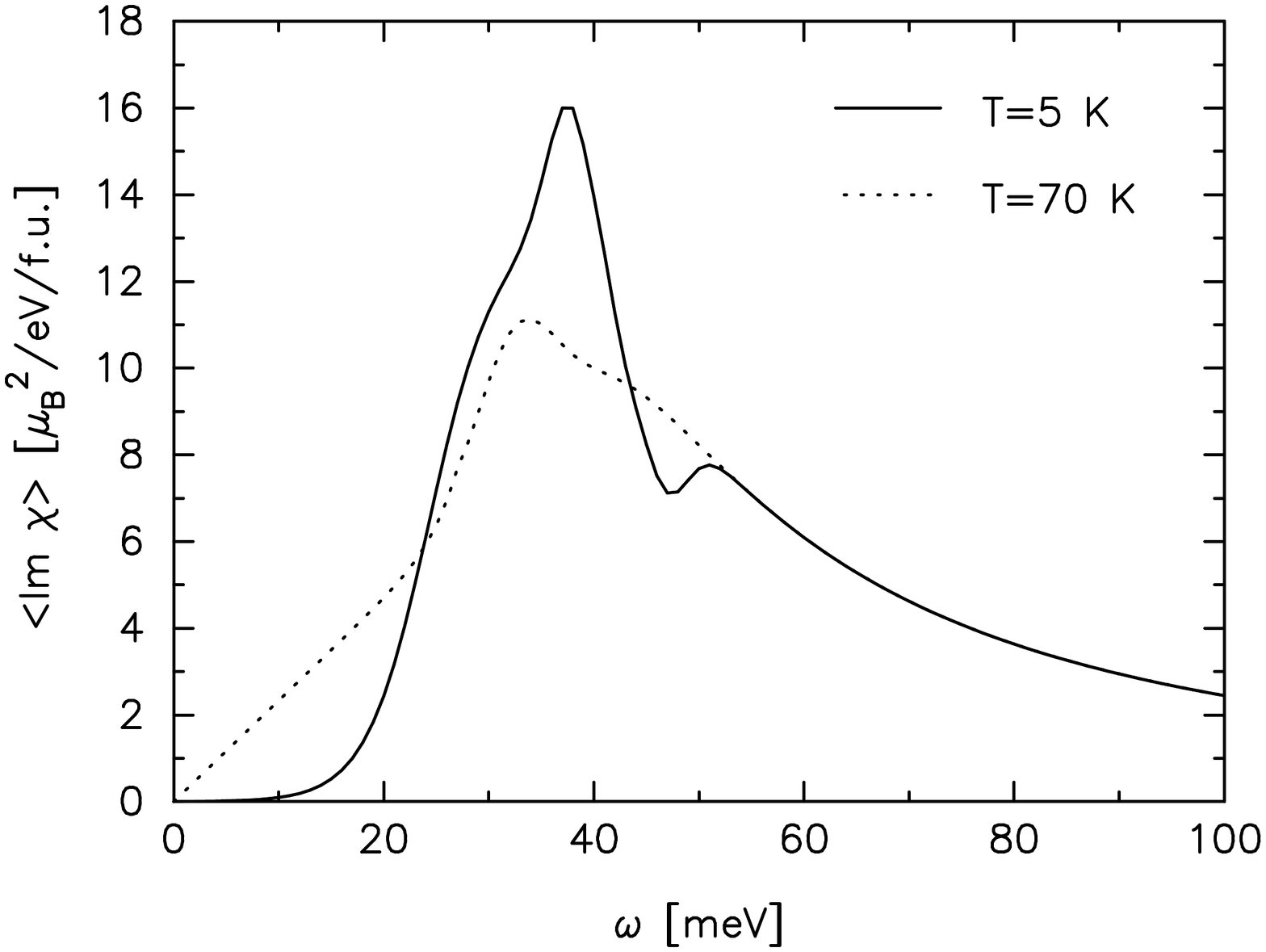}
\end{center}
\caption{Momentum-integrated spin excitation spectra at $T=5$ K and
$T=70$ K according to the fitting formulae.} \label{fig:localimchi}
\end{figure}

\begin{figure}[htb]
\begin{center}
\includegraphics[width=0.6 \columnwidth, angle=0]{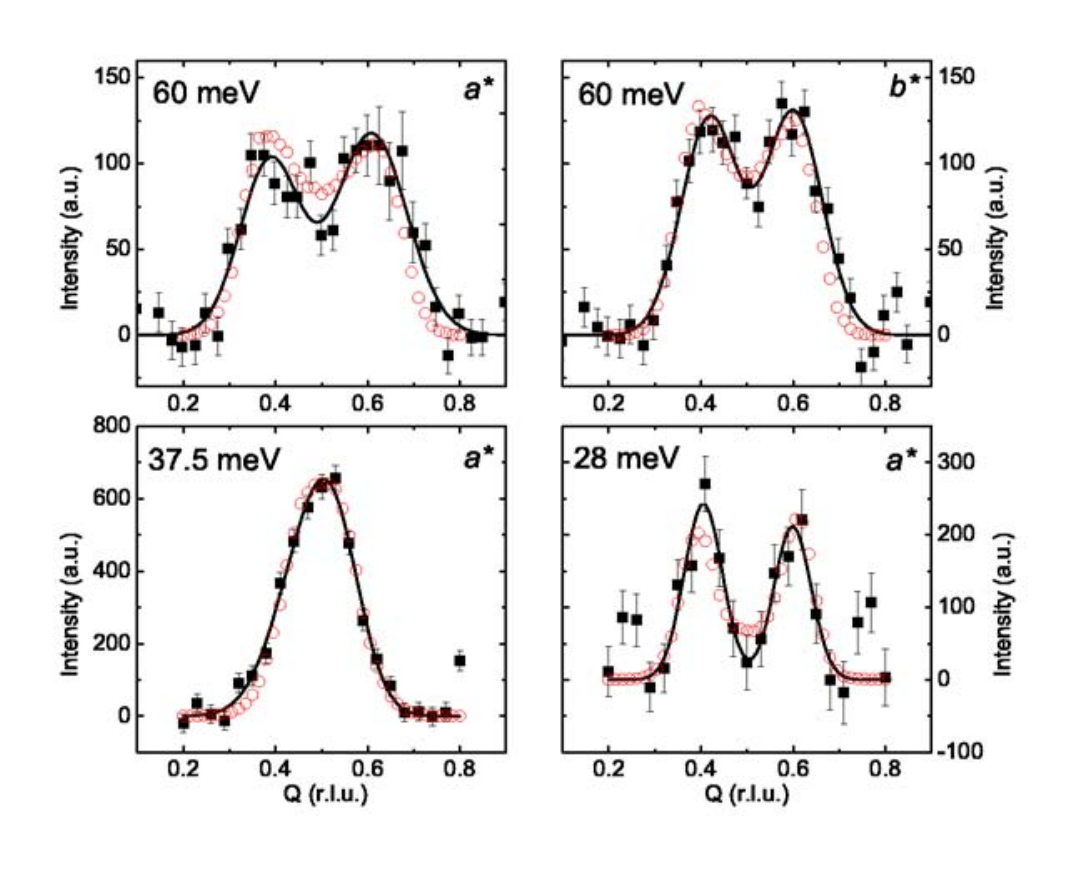}
\end{center}
\caption{Comparison of typical constant-energy scans at $T=5$~K (black squares) with our fit formula convoluted with the instrumental resolution function (open cirles). The black lines are guides to the eye for the measured data points. High- and low-energy scans were measured at different $k_f$, thus the intensities are not comparable.} \label{fig:fitExamples}
\end{figure}

\begin{figure}[htb]
\begin{center}
\includegraphics[width= 0.6 \columnwidth, angle=270]{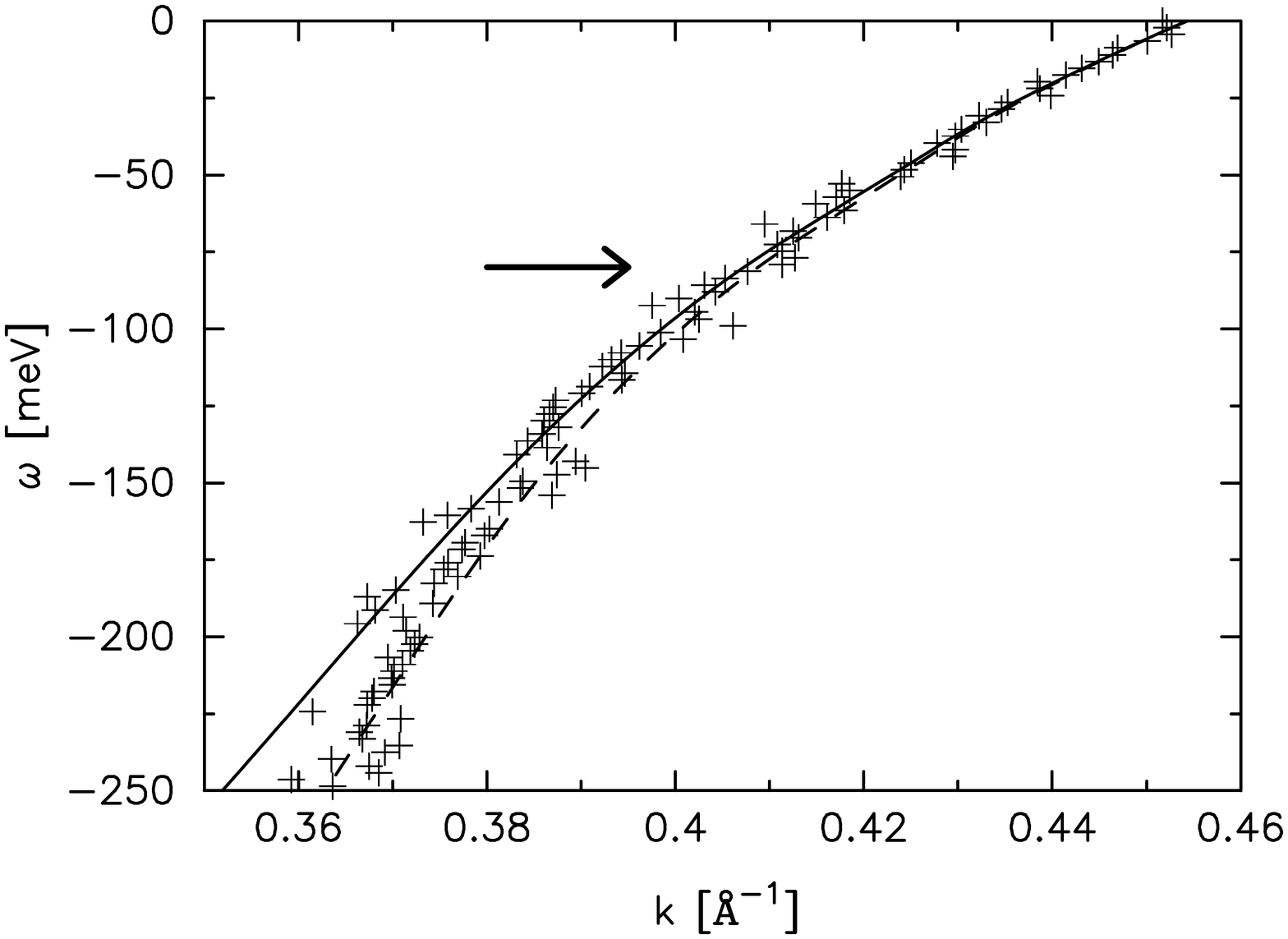}
\end{center}
\caption{Comparison of the nodal dispersion measured by ARPES (crosses) and evaluated theoretically for unrenormalized Fermi velocities of 4 eV\AA\ (solid line) and 5 eV\AA\ (dashed line).
}
\end{figure}

\end{document}